\documentclass[preprint,prd,noshowpacs,nofootinbib]{revtex4}

\usepackage{color} 
\usepackage{amssymb}
\usepackage{amsmath}
\usepackage{graphicx} 
\usepackage{float}
\usepackage{braket}     
\usepackage{pdfpages}
\usepackage{epstopdf}
\usepackage[utf8]{inputenc}

\begin{document}

\title{ Momentum Gauge Fields and Non-Commutative Space-Time }

\author{Eduardo Guendelman}
\email{guendel@bgu.ac.il}  \email{guendelman@fias.uni-frankfurt.de}
\affiliation{Department of Physics, Ben-Gurion University of the Negev, Beer-Sheva, Israel \vspace{0.15in} \\
Frankfurt Institute for Advanced Studies, Giersch Science Center, Campus Riedberg, Frankfurt am Main, Germany \vspace{0.15in} \\
Bahamas Advanced Studies Institute and Conferences, 4A Ocean Heights, Hill View Circle, Stella Maris, Long Island, The Bahamas}

\author{Douglas Singleton}
\email{dougs@mail.fresnostate.edu}
\affiliation{Physics Department, California State University Fresno, Fresno, CA 93740}

\date{\today}

\begin{abstract}
 In this work we present a gauge principle that starts with the momentum space representation of the position operator (${\hat x}_i = i \hbar \frac{\partial}{\partial p_i}$) rather than starting with the position space representation of the momentum operator (${\hat p}_i = -i \hbar \frac{\partial}{\partial x_i}$). This extension of the gauge principle can be seen as a dynamical version of Born's reciprocity theory which exchanges position and momentum. We discuss some simple examples with this new type of gauge theory: (i) analog solutions from ordinary gauge theory in this momentum gauge theory, (ii) Landau levels using momentum gauge fields, (iii) the emergence of non-commutative space-times from the momentum gauge fields. We find that the non-commutative space-time parameter can be momentum dependent, and one can construct a model where space-time is commutative at low momentum but becomes non-commutative at high momentum. 
\end{abstract}

\maketitle

\section{Gauge theory in momentum space}
Gauge theories have been one of the central ideas of theoretical physics in the past hundred years \cite{straumann,abers}. The Standard Model of particle physics, which describes all known non-gravitational interactions, is a gauge theory \cite{sm,RelativisticQuantumMechanicsandRelatedTopics} and general relativity can be viewed as a gauge theory \cite{lasenby}. It is very important to emphasize the central role of Professor Steven Weinberg in the development and applications of the gauge principle in the construction of what we call now the Standard Model. 
Professor Weinberg was also very active into the issue of extending the Standard Model, exploring the ideas of axions \cite{w-axion}, supersymmetry \cite{w-susy}, string theory  and cosmological issues \cite{w-cosmo}. Much of Professor Weinberg's work dealt with symmetries in physics and their applications. In this work we present a new  extension of the gauge symmetry principle, by extending the usual gauge symmetry to momentum space.

In the standard formulation of a gauge theory one starts with a space-time dependent matter field $\Psi (x)$ which satisfies some matter field equation ({\it e.g.} Schr{\"o}dinger equation, Klein-Gordon equation, Dirac equation) and requires that this matter field satisfy a local phase symmetry of the form $\Psi (x) \to e^{- i \lambda (x)} \Psi (x)$. The gauge function, $\lambda (x)$, can depend on space and time. Along with this local phase symmetry of the matter field, one needs to introduce the kinetic momentum/gauge covariant derivative $p_i \to p_i - e A_i (x)$ or $\frac{\partial}{\partial x_i} \to  \frac{\partial}{\partial x_i} -  i e A_i (x)$, where the vector potential obeys $A_i (x) \to A_i (x) - \frac{1}{e} \frac{\partial \lambda (x)}{\partial x_i}$. This standard construction is done in position space: the matter field, $\Psi$ is a function of position, the momentum operator is given as a derivative of position ($p_i = -i \frac{\partial}{\partial x_i}$ and we take $\hbar =1$), and the vector potential and gauge function are functions of space and time coordinates. 

However, quantum mechanics can be carried out in momentum space as well with the matter field being a function of momentum, $\Psi (p)$, and the position operator being given by $x_i = i \frac{\partial}{\partial p_i}$. In this construction the momentum operator is just multiplication by $p_i$ just as the position operator in position space is
multiplication by $x_i$. The momentum space gauge transformation of the matter field should be 
\begin{equation}
    \label{mom-gauge}
    \Psi (p) \to e^{- i \eta (p)} \Psi (p) ~.
\end{equation}
The equivalent of the generalized position/gauge covariant derivative is
\begin{equation}
    \label{mom-derivative}
    x_i \to x_i - g C_i (p) ~~~~{\rm or}~~~~ \frac{\partial}{\partial p_i} \to  \frac{\partial}{\partial p_i} +  i g C_i (p).
\end{equation}
We have used $x_i = i \frac{\partial}{\partial p_i}$, $g$ is some momentum-space coupling, and $C_i (p)$ is a momentum-space gauge function which must satisfy 
\begin{equation}
    \label{mom-gauge-field}
     C_i (p) \to C_i (p) + \frac{1}{g} \frac{\partial \eta (p)}{\partial p_i}.
\end{equation}
Finally one can construct a momentum-space field strength tensor which is invariant under just \eqref{mom-gauge-field}, namely
\begin{equation}
    \label{guv}
    G_{ij} = \frac{\partial C_i}{\partial p_j} - \frac{\partial C_j}{\partial p_i}.
\end{equation}
This is the $p_ip_j$ component of the momentum gauge field, field strength tensor. It is the analog of $x_ix_j$ component of the standard gauge field, field strength tensor $F_{ij} = \frac{\partial A_i}{\partial x_j} - \frac{\partial A_j}{\partial x_i}$. The 4-vector version of the standard gauge potential and field strength tensor are $A_i \to A_\mu$ and $F_{ij} \to F_{\mu \nu}$. One needs to make a similar 4-vector/4-tensor extension for the momentum gauge field and associated field strength tensor via  $C_i (p) \to C_{\mu} (p)$ and $G_{ij} \to G_{\mu \nu}$. The momentum generalized gauge field and field strength tensor are reminiscent of the Berry connection and Berry curvature \cite{berry1984}, where the Berry connection/Berry ``gauge" field is the function of some parameter, that is not necessarily position. Here $C_i (p)$ and $G_{ij} (p)$ are Berry connections and Berry curvatures that are specifically functions of momentum. 

One can ask about the units of the momentum coupling, $g$ and momentum gauge field, $C_\mu$, relative to the standard coupling, $e$, and standard gauge field, $A_\mu$. From $p_i \to p_i - e A_i (x)$ and $x_i \to x_i - g C_i (p)$ one sees that $e A_i (x)$ has units of momentum while $g C_i (p)$ has units of position. This leaves two options for the units of $g$ and $C_\mu$. First one can choose for $g$ to have the same units as $e$ and then the units of $C_\mu$ would be the units of $A_\mu$ multiplied by $\frac{[position]}{[momentum]} = \frac{[time]}{[mass]}$. Second one can choose for $C_\mu$ and $A_\mu$ to have the same units and in this case the units of $g$ would be the units of $e$ again multiplied by the same factor $\frac{[position]}{[momentum]} = \frac{[time]}{[mass]}$. 

One can ask if there is some deeper connection or condition between the standard coupling $e$  and momentum coupling $g$, perhaps something like the Dirac quantization condition \cite{dirac} between electric and magnetic charge. One idea might be to take the option above where $e$ and $g$ have the same units and then via Born reciprocity require the couplings to be exchangeable {\it i.e.} $e \leftrightarrow g$. We leave this question for future work. 

The above discussion shows that one can easily construct a momentum-space analog of the canonical position-space gauge procedure. There are two questions this raises: (i) What physical use/significance would this momentum gauge field construction have? (ii) Why is this momentum gauge field construction not as common as the standard gauge field construction? The first question will be addressed in following sections, but here we will address the second question. The answer may lie in the asymmetric way in which the momentum and position operators appear the in simplest, free particle Hamiltonian. For a  non-relativistic object of mass $m$ this Hamiltonian is
\begin{equation}
    \label{H}
H = \frac{1}{2m}(p_1^2 +p_2^2 + p_3^2)= - \frac{\hbar ^2}{2m} \left( \frac{\partial ^2}{\partial x^2_1} + \frac{\partial ^2}{\partial x^2_2}+ \frac{\partial ^2}{\partial x^2_3} \right) = - \frac{\hbar ^2}{2m} \nabla ^2 ~.
\end{equation}
The Hamiltonian in \eqref{H} is suited for the covariant derivative $p_i \to p_i - e A_i (x)$ or $\frac{\partial}{\partial x_i} \to  \frac{\partial}{\partial x_i} -  i e A_i (x)$ but there is no room, nor use for the momentum space version in \eqref{mom-derivative}. However, a more symmetric starting point would be to consider the non-relativistic simple harmonic oscillator Hamiltonian
\begin{equation}
    \label{H-sho}
H = \frac{1}{2m} (p_1^2 + p_2^2 + p_3^2) + \frac{m \omega ^2}{2} (x_1^2 + x_2^2 +x_3^2 )  \to  \frac{1}{2}(p_1^2+ p_2^2 + p_3^2) + \frac{1}{2}(x_1^2+x_2^2+x_3^2) ~.
\end{equation}
In the last step we have chosen the mass and frequency of the oscillator as $m=1$ and $\omega=1$. Looking at the last form in \eqref{H-sho} one sees a symmetry between momentum and position of $p_i \leftrightarrow x_i$. This symmetry provides an argument to have the gauge principle apply not only the momentum via $p_i \to p_i - e A_i$, but also to the position via $x_i \to x_i -g C_i$. One can argue for the naturalness of $\frac{p_i^2}{2m} + \frac{m \omega ^2}{2} x_i^2$ over just $\frac{p_i^2}{2m}$ by pointing to the quantum field theory (QFT) vacuum, which can be viewed as a collection of harmonic oscillators \cite{zee}, so that having both the momentum and position terms in the Hamiltonian is more natural than have only momentum or only position.

In the above we have exchanged the roles of the position and momentum operators in the usual construction of a gauge theory. The momentum gauge fields provides for  a dynamical model of Born reciprocity \cite{born1949} which exchanged the position and momentum operators as ${\hat x} \to {\hat p}$ and ${\hat p} \to - {\hat x}$, which would imply that for every ordinary gauge field there should be a corresponding momentum gauge field. For example the Hamiltonian in \eqref{H-sho} is invariant under this swap of position and momentum in units where $m=1$ and $\omega =1$. The minus sign in the momentum to position transformation keep the standard form of the position-momentum commutator under this change {\it i.e.} $[{\hat x}, {\hat p}] = i \hbar$ is invariant under ${\hat x} \to {\hat p}$ and ${\hat p} \to - {\hat x}$. The Hamiltonian in \eqref{H-sho} is also invariant under the transformation ${\hat x} \to {\hat p}$ and ${\hat p} \to {\hat x}$, but this would then change the sign of the position-momentum commutator $[{\hat x}, {\hat p}] = - i \hbar$. Nevertheless this would still lead to the same uncertainty principle since $\Delta x \Delta p = \frac{1}{2} \left|  \langle [{\hat x}, {\hat p}] \rangle \right|$. The main difference between the model we lay out here and that in reference \cite{born1949} is that we have included a momentum gauge field \eqref{mom-gauge-field} and momentum field strength tensor \eqref{guv}, thus making possible  a dynamical model of Born reciprocity.  

\section{Connection to non-commutative space-time}

\subsection{Constant non-commutativity parameter}

In this subsection we point out the connection of the above momentum gauge theory with non-commutative geometry, by which we mean coordinates obeying 
\begin{equation}
    \label{nc-geo}
    [x_i, x_j] = i \Theta _{ij} ~,
\end{equation}
where $\Theta _{ij}$ is an anti-symmetric, {\it constant} rank-2 tensor. A review of non-commutative field theory can be found in \cite{Noncommutativefieldtheory}, and interesting applications of non-commutative geometry to modifications on the hydrogen atom spectrum can be found in \cite{NoncommutativeHydrogenatom}. There are also work that looks at how non-commutative geometry may cure the singularities found in black holes and other solutions in general relativity \cite{piero-2009}. The construction from the previous section leads exactly to this kind of non-commutativity between the coordinates. We begin with equation \eqref{mom-derivative} and define a generalized, gauge invariant coordinate $X_i = x_i -g C_i (p) =  i \partial_{p_i} - g C_i (p)$. In its first form this looks like coordinate translation by $g C_i (p)$. Calculating the commutator of $X_i$ and $X_j$ gives
\begin{equation}
    \label{x-com}
    [X_i, X_j] = i g G_{ij} ~,
\end{equation}
with the momentum-space field strength $G_{ij}$ defined in \eqref{guv}. Equation \eqref{x-com} is of the form \eqref{nc-geo} with $\Theta _{ij} = g G_{ij}$.

The result in \eqref{x-com}  is reminiscent of the non-commutativity of the covariant derivative for regular, minimally coupled fields,  $\pi _i = p_i - e A_i (x) = -i \partial_{x_i} - e A_i (x)$. Calculating the commutator of $\pi _i$ with $\pi _j$ gives
\begin{equation}
    \label{b-field}
    [\pi_i, \pi_j ] = i e F_{ij} = i e \epsilon _{ijk} B^k ~,
\end{equation}
where $B^k = \frac{1}{2} \epsilon^{kij} (\partial _{x_i} A_j - \partial _{x_j} A_i) = \frac{1}{2} \epsilon^{kij} F_{ij}$ is the regular magnetic field. Comparing \eqref{x-com} with \eqref{b-field} one can define a momentum gauge field ``magnetic field" as ${\cal B}^k = \frac{1}{2} \epsilon^{kij} (\partial _{p_i} C_j - \partial _{p_j} C_i) = \frac{1}{2} \epsilon^{kij}G_{ij}$. This in turn defines the non-commutation parameter of the spatial coordinates on the right hand side of \eqref{x-com} to be constant only if the momentum ``magnetic" field is constant. 

One can easily arrange for such a constant ``magnetic" field solution via
\begin{equation}
\label{constantmomentummagneticfield}
C^0 = 0 ~~~,~~~  C^{i} =\frac{1}{2} \epsilon^{ijk} {\cal B}^{j}p^{k}
\end{equation}
with $ {\cal B}^{j} $ being a constant. Taking the curl of \eqref{constantmomentummagneticfield}, using momentum derivatives, and doing index gymnastics yields $\epsilon ^{lmi} \partial ^{p^m} C^i  = {\cal B}^l$ -- one gets a constant ``magnetic" field.  This gives a constant non-commutative tensor $\Theta_{ij} = g G_{ij} = g \epsilon _{ijk} {\cal B}^k$ {\it i.e.} in this way one recovers a constant non-commutative parameter which is the most common assumption in the literature \cite{Noncommutativefieldtheory,NoncommutativeHydrogenatom}. 

A fully 4-vector version of the spatial coordinate non-commutativity in \eqref{nc-geo} is accomplished by promoting the 3-Latin indices to 4-Greek indices giving
\begin{equation}
    \label{nc-geo-1}
    [x_\mu, x_\nu] = i \Theta _{\mu \nu} ~,
\end{equation}
where $\Theta _{\mu \nu}$ is an anti-symmetric 4-tensor. In conjunction with \eqref{nc-geo-1} the 4-tensor version of \eqref{x-com} becomes 
\begin{equation}
    \label{x-com-1}
    [X_\mu, X_\nu] = i g G_{\mu \nu} ~,
\end{equation}
In order to get a constant $\Theta _{\mu \nu}$ for a component with one space index ({\it e.g.} $\mu =i$) and one time index ({\it i.e.} $\nu =0$) we need to have a constant momentum gauge field, ``electric" field.  This is accomplished by selecting the momentum gauge field as 
\begin{equation}
\label{constantmomentumelectricfield}
C^0 = -{\cal E} ^{j} p^{j} ~~~~;~~~~  C^{j} = 0
\end{equation}
The momentum gauge ``electric" field  is given by $G_{0i}= \partial _{p^0} C_i - \partial _{p^i} C_0 = {\cal E}^i$ which is the sought after constant momentum gauge field ``electric" field. Using equations \eqref{nc-geo-1} and \eqref{x-com-1} this gives the connection between the non-commutativity parameter and momentum gauge field electric field of $\Theta_{0i} = g G_{0i} = g {\cal E}_i$.

\subsection{Variable non-commutativity parameter}

In the previous subsection we looked at momentum gauge field configuration with constant ``magnetic" and constant ``electric fields" in equations  \eqref{constantmomentummagneticfield} and \eqref{constantmomentumelectricfield} respectively. In this subsection we examine momentum gauge field configurations which are variable. These variable momentum gauge fields then imply a varying of the non-commutativity parameter via the connection $\Theta _{uv} \propto G_{\mu \nu}$. 

We first write down two common, ordinary gauge field solutions which have gauge fields that vary with space and time and then construct the varying momentum gauge field analogs. The two ordinary gauge field solutions we consider are a plane wave and a static points charge. The Lagrange density for standard gauge fields is ${\cal L}_F= -\frac{1}{4} F_{\mu \nu } F^{\mu \nu}$ with $F^{\mu \nu} =\partial ^{x^\mu} A^\nu - \partial ^{x^\nu} A^\mu$. The equations of motion from ${\cal L}_F$ are
\begin{equation}
\label{eom-gauge}
\partial_{x_\mu} (\partial ^{x_\mu} A^\nu - \partial ^{x_\nu} A^\mu) = 4 \pi J^\nu (x) \to \partial_{x_\mu} \partial ^{x_\mu} A^\nu  = 4 \pi J^\nu (x) \to \Box _x  A^\nu = 4 \pi J^\nu (x)~,
\end{equation}
with $J^\nu (x)$ being a conserved 4-current coming from some matter source, and $\Box _x$ is the d'Alembertian with respect to the time-position coordinates. In the last line we have taken the Lorenz gauge $\partial _{x_\mu} A^{x_\mu} = 0$.  Let us look at two common solutions to \eqref{eom-gauge}: the plane wave vacuum solution and the point charge solution.
\begin{itemize}
    \item In vacuum ($J^\nu =0$) \eqref{eom-gauge} has the solution $A^\nu \propto e^{i(px - E t)} \varepsilon^\nu \delta (p^2 - E^2 /c^2)$ where the $\delta$-function enforces the mass shell condition $\frac{E ^2}{p^2} =c^2$ and $\varepsilon ^\nu$ is the polarization vector. 
    \item For a point charge at rest one has the current $J^\nu = (q \delta^3 (r), 0, 0, 0)$, which has the solution $A^0 = \frac{q}{r}$ and ${\vec A} = 0$, since $\nabla _x ^2 \left( \frac{1}{r}\right) = 4 \pi \delta (r)$ .
\end{itemize}

We now examine how the above plays out for the momentum  gauge fields. The momentum gauge field Lagrange density is ${\cal L}_G= -\frac{1}{4} G_{\mu \nu } G^{\mu \nu}$ with $G^{\mu \nu} = \partial ^{p_\mu} C^\nu - \partial ^{p_\nu} C^\mu$. The equations of motions that follow from this Lagrange density are
\begin{equation}
\label{eom-gauge-2}
\partial_{p_\mu} (\partial ^{p_\mu} C^\nu - \partial ^{p_\nu} C^\mu) = 4 \pi {\cal J}^\nu (p) \to \partial_{p_\mu} \partial ^{p_\mu} C^\nu  = 4 \pi {\cal J}^\nu (p) \to  \Box _p C^\nu = 4 \pi {\cal J}^\nu (p) ~,
\end{equation}
with ${\cal J}^\nu (p)$ being a 4-current matter source that is a function of $p$, and $\Box _p $ is the d'Alembertian with respect to energy-momentum. In the last expression we use the momentum space equivalent of the Lorenz gauge $\partial _{p_\mu} C^{p_\mu} = 0$.  The current conservation in momentum space reads   $\partial _{p_\mu} {\cal J}^{p_\mu} = 0$,

We now repeat the two types of solutions listed above for the standard gauge theory, but for the momentum gauge theory. 
\begin{itemize}
    \item In vacuum (${\cal J}^\nu =0$) \eqref{eom-gauge-2} has  solution $C^\nu \propto e^{i(px - E t)} \varepsilon ^\nu \delta (x^2 - c^2 t^2)$ where the $\delta$-function enforces the light-cone condition $\frac{x ^2}{t^2} = c^2$, and $\varepsilon ^\nu$ is the polarization vector. 
    
    \item The momentum gauge equivalent of the charge at rest is given by ${\cal J}^\nu = (g \delta^3 (p), 0, 0, 0)$, with $C^0 = \frac{g}{p}$ and ${\vec C} = 0$ since  $\nabla _p ^2 \left( \frac{1}{p}\right) = 4 \pi \delta (p)$.
\end{itemize}
Notice that the point source in momentum space, that is ${\cal J}^\nu = (g \delta^3 (p), 0, 0, 0)$ is a totally homogeneous solution in coordinate space, since it is concentrated at zero momentum, which means indeed the assumption of a totally homogeneous state.  More generally, it is interesting to observe that any current of the form ${\cal J}^\nu = (f (p), 0, 0, 0)$ , with ${\cal J}^0  = f ({\vec p}) $ being $p^0$ independent, will satisfy the current conservation law of  $\partial _{p_\mu} {\cal J}^{\mu} = 0$. Doing a Fourier transformation on this to coordinate space yields $x_\mu {\tilde {\cal  J}}^\mu = 0$, where ${\tilde {\cal  J}}^\mu$ is the Fourier transformation of ${\cal  J}^\mu$. The equivalent statements for a a regular 4-source would be $J^\nu = (f({\vec x}) , 0, 0, 0)$, which satisfies the conservation law  $\partial _{x_\mu} J^{\mu} = 0$ or Fourier transforming to momentum space $p_\mu {\tilde   J}^\mu = 0$.

One can construct other conserved current sources for momentum gauge fields that satisfy  $x_\mu {\cal {\tilde J}}^\mu = 0$. Starting with any 4-vector $V^\mu$, we construct
${\cal {\tilde J}}^\mu = V^\mu- x^\mu V^\nu  x_\nu  /x ^2$ which is easily seen to satisfy $x_\mu {\cal {\tilde J}}^\mu =0$.

As a final comment the equation of motion for, $A_\mu$, given in \eqref{eom-gauge} leads to a propagator in momentum space that is proportion to $\propto \frac{1}{p^2 + i \epsilon}$. The imaginary infinitesimal term $i \epsilon$ is a convergence factor to deal with the divergence as $p \to 0$. In turn the momentum gauge field equation of motion, $C_\mu$, given in \eqref{eom-gauge-2}, leads to a position space propagator proportional to $\propto \frac{1}{x^2}$. Here we have not inserted a factor of $i \epsilon$. This is because the limits $p \to 0$ and $x \to 0$ are physically different. The $p \to 0$ limit is the infrared/low energy limit which is dealt with by inserting a convergence factor of $i \epsilon$ which is taken to zero at the end. The $x \to 0$ limit is the ultraviolet/high energy limit which is dealt with using the renormalization procedure if the theory turns out to be renormalizable, or by introducing a cut off if the theory turns out to be non renormalizable. 

\section{Generalized Landau Levels}

In this section we work on the case of generalized Landau levels with a particle of mass $m$ in a constant ordinary magnetic field and constant momentum ``magnetic" field. We take both the ordinary and momentum magnetic field to point in the $3/z$-direction.  We want to take these magnetic fields and minimally couple them to the free particle in equation  \eqref{H-sho}. Applying minimal coupling for both coordinate gauge fields and momentum gauge fields, leads to  $p_i \to p_i - e A_i$, and $x_i \to x_i -g C_i$. Having a constant, ordinary magnetic field and a constant, momentum magnetic fields in the $3/z$-direction can be obtained in the symmetric gauge with $A_1$ and $A_2$ given by,
\begin{equation}
\label{constantcoordinatemagneticfieldincoordinate12}
A_0 = 0 ~~,~~  A_1 = -\frac{1}{2}  B y ~~,~~   A_2 = \frac{1}{2}  B x ~,
\end{equation}
and with $C_1$ and $C_2$ also in the symmetric gauge given by,
\begin{equation}
\label{constantmomentummagneticfieldincoordinate12}
C_0 = 0 ~~,~~  C_1 = -\frac{1}{2} {\cal B} p_y  ~~,~~   C_2 =  \frac{1}{2} {\cal B} p_x~,
\end{equation}
The constant values of the ordinary magnetic field and momentum magnetic field from \eqref{constantcoordinatemagneticfieldincoordinate12} and \eqref{constantmomentummagneticfieldincoordinate12} are $B$ and ${\cal B}$ respectively. 

So the equation of motion for the double gauged harmonic oscillator reads,
\begin{eqnarray}
    \label{H-shogauged}
H &=&  \frac{1}{2m} \left(p_x + \frac{e B y}{2}\right)^2 +\frac{1}{2m} \left( p_y - \frac{e  B x}{2} \right)^2 \nonumber \\
&+& \frac{m \omega ^2}{2} \left( x + \frac {g{\cal B} p_y}{2} \right)^2 +\frac{m \omega ^2}{2} \left( y - \frac{g {\cal B} p_x}{2} \right)^2 
+ \frac{p_z ^2}{2m} + \frac{m \omega ^2}{2} z^2  
\end{eqnarray}
or (we drop the part of the Hamiltonian associated with the kinetic energy and harmonic oscillator in the $z$-direction)
\begin{equation}
    \label{H-shogauged expandes}
H = \left(1+\frac{(gm \omega {\cal B} )^2}{4}\right) \left(\frac{p_x^2}{2m} +\frac{p_y^2}{2m}\right) +  \left(1+\frac{(e B )^2}{4m^2\omega ^2}\right) \frac{m \omega ^2}{2}( x^2 +y^2) +L_z(-g_1 B+g_2
{\cal B})~.
\end{equation}
Here $L_z =x p_y - y p_x$, this the angular momentum in the $z$-direction. and $g_1= \frac{e}{2m}$ and  $g_2 = \frac{gm\omega ^2}{2}$ are the coupling strengths of the angular momentum to the coordinate magnetic field $B$ and the momentum magnetic field ${\cal B}$ respectively. 

 The above results can be compared with the formulation of non-commutative quantum mechanics \cite{Noncommutativequantummechanics} for the case of an harmonic oscillator potential, and the results agree with those in  \cite{Noncommutativequantummechanics}, if the identification of the non-commutative parameter is made according to expression  \eqref{x-com}. 
 
 The coupling between $B$ and $L_z$ is exactly what one has from the standard analysis of Landau levels. The coupling between $L_z$ and ${\cal B}$ is a new feature arising from the momentum gauge fields, but the two coupling terms to $L_z$ have a dual symmetry between the regular magnetic field, $B$, and momentum gauge ``magnetic" field, ${\cal B}$. 
 
The first term in \eqref{H-shogauged expandes} shows that the system has now developed a new, effective mass  given by 
 \begin{equation}
     \label{m-eff}
     m_{eff} = \frac{m}{1+\frac{(gm \omega {\cal B} )^2}{4}}~.
 \end{equation}
The effective mass depends on the momentum ``magnetic" field and is always less than $m$ {\it i.e.} $m_{eff} < m$. In addition the second terms in \eqref{H-shogauged expandes} implies a new effective frequency. Taking into account the effective mass in \eqref{m-eff} to write this second term in the form $\frac{m_{eff} \omega _{eff}  ^2}{2}$ gives a new effective frequency of
 \begin{equation}
     \label{weff}
     \omega _{eff} = \omega \sqrt{\left( 1+\frac{(gm \omega {\cal B} )^2}{4}\right) \left(1 + \frac{e^2 B^2}{4 m^2 \omega^2}\right)} ~.
 \end{equation}
Note that in the effective frequency and the effective mass above, both momentum ({\it i.e.} ${\cal B}$) and coordinate magnetic ({\it i.e.} $B$) fields  contribute. 

One can define an effective magnetic field as
 \begin{equation}
     \label{Beff} 
 B_{eff} = \frac{-g_1 B+g_2{\cal B}}{\sqrt{g^2_1+g^2_2}} ~,
 \end{equation}
 so that the coupling of the $z$-component of angular momentum to the two magnetic fields, ${\cal B}$ and $B$ in \eqref{H-shogauged expandes} can be written as $\sqrt{g^2_1+g^2_2}B_{eff}L_z $. One can also define a generalized magnetic field orthogonal to $B_{eff}$ via 
 \begin{equation}
     \label{Bno} 
 B_{nc} = \frac{g_1{\cal B}+ g_2 B}{\sqrt{g^2_1+g^2_2}} ~.
 \end{equation}
 The subscripts $nc$ stand for ``non-coupling" since $B_{nc}$, unlike $B_{eff}$, does not couple to $L_z$.

 The definition of the two generalized magnetic fields in \eqref{Beff} and \eqref{Bno} is mathematically identical to the definition of the $Z^0$ boson and photon in the Standard Model \cite{abers,sm,RelativisticQuantumMechanicsandRelatedTopics}. Further from \eqref{Bno} , \eqref{Beff} we can define
 an analog of the ``Weinberg angle" via the definition 
 \begin{equation}
     \label{mix} 
cos (\theta_{mixing} ) = \frac{g_1}{\sqrt{g^2_1+g^2_2}} ~.
 \end{equation}
 
Putting all of the above together the total Hamiltonian is then,
 \begin{equation}
     \label{TOTALH}
    H = \frac{1}{2 m_{eff}}(p_x^2+p_y^2)+\frac{1}{2}\omega _{eff}^2 m_{eff}(x^2+y^2)+ \sqrt{g^2_1+g^2_2}B_{eff}L_z
 \end{equation}
Note that $B_{eff}$ couples to the angular momentum, while $B_{nc}$ does not. This is similar to the Standard model where the $Z^0$ has a mass term while the photon remains massless. 
 
 Following \cite{abers-2} one can define creation/annihilation operators in terms of $p_x, p_y$ and $x, y$ as
 \begin{eqnarray}
     \label{cre-ann}
      x  &=& \sqrt{\frac{\hbar}{2\omega_{eff} m_{eff}}} \left( a_1 + a^\dagger _1 \right) ~~~;~~~
      y  = \sqrt{\frac{\hbar}{2\omega_{eff} m_{eff}}} \left( a_2 + a^\dagger _2 \right) \nonumber \\
      ~~~{\rm and}~~~ \\
      p_x &=& i \sqrt{\frac{\hbar \omega_{eff} m_{eff}}{2}} \left( a^\dagger _1 - a^\dagger _1 \right) ~~~;~~~
      p_y = i \sqrt{\frac{\hbar \omega_{eff} m_{eff}}{2}} \left( a^\dagger _2 - a^\dagger _2 \right) \nonumber~.
 \end{eqnarray}
 The creation and annihilation operators obey the usual relationship $[a_i, a^\dagger _j] = \delta _{ij}$. With these definitions we find $L_z = x p_y - y p_x = i \hbar (a_1 a_2 ^\dagger - a_2 a_1 ^\dagger)$ and the Hamiltonian in \eqref{TOTALH} becomes $H = \hbar \omega_{eff} (a^\dagger _1 a_1 + a^\dagger _2 a_2 +1 ) + i \hbar \sqrt{g^2_1+g^2_2}B_{eff} (a_1 a_2 ^\dagger - a_2 a_1 ^\dagger)$. The first two terms can be seen to be the normal 2D harmonic oscillator. The third term looks a like a coupling between the generalized magnetic field and the angular momentum in the $z$ direction. %Notice that unlike the standard Landau Levels, the strength of the angular momentum coupling to the magnetic field is not correlated to the harmonic oscillator frequency.

\section{Momentum dependent non-commutativity parameter}

In this section we examine two simple examples where the non-commutativity parameter, $\Theta _{\mu \nu}$, is not a constant but depends on the momentum. Recently, other authors \cite{momentumdependenttheta} have considered momentum dependent non-commutative parameters. However, in this work the inspiration is quite different as it exploits some geometry in momentum space. Also the non-commutativity parameter in \cite{momentumdependenttheta} depends on both momentum and position, while our in our construction below the non-commutativity parameter depends only on momentum, which is closer to the energy-momentum dependence of masses and couplings in QFT that one finds from the renormalization group. 

The examples we choose are the momentum gauge field version of a capacitor and solenoid, with the momentum gauge fields being piece-wise constant in different momentum ranges, leading to different, $\Theta_{\mu \nu}$'s in these different ranges. 

\subsection{Capacitor-type momentum electric field configuration}

The standard, infinite parallel plate capacitor has a 4-current source of  
\begin{equation}
\label{e-source}
    J^\nu = (f (z), 0, 0, 0) ~~ {\rm with} ~~ f (z)  =\sigma [\delta(z + a) - \delta(z - a)]
\end{equation}
This source represents two infinite planes of surface charge $\pm \sigma$ placed perpendicular to the $z$-axis at $z= \mp a$. This source gives an electric field of 
\begin{equation}
\label{capacitor}
    E_z = 4 \pi \sigma ~~ {\rm for} ~~ - a \le z \le a ~~~ {\rm and} ~~~ E_z = 0 ~~ {\rm for}~~  |a| \le |z| ~, 
\end{equation}
{\it i.e.} non-zero between the planes and zero outside the planes. 

The momentum gauge field analog of this standard capacitor system has a constant momentum ``electric" field similar to that in equation \eqref{constantmomentumelectricfield}, but it should be restricted in momentum rather than position as is the case in equation \eqref{capacitor}. Actually for the momentum gauge field system we want the inverse of the above standard capacitor -- we want the momentum ``electric" field to be zero between the planes ({\it i.e} at small momentum) and non-zero outside the planes ({\it i.e} at large momentum). The capacitor-like configuration for the momentum gauge fields that we want has a 4-current source of
\begin{equation}
    \label{e-source-m}
    {\cal J}^\nu = (f (p), 0, 0, 0) ~~ {\rm with}~~ f (p)  =\Sigma [\delta(p_z + p_a) + \delta(p_z - p_a)]. 
\end{equation}

The planes are symmetrically placed at $p_z = \pm p_a$ and, in contrast to the sources for the standard capacitor in \eqref{e-source}, the momentum planes now have the {\bf same} ``surface charge", $\Sigma$. This same ``surface charge" set up leads to a momentum ``electric" field in the $p_z$ direction given by 
\begin{eqnarray}
    \label{m-capacitor}
    {\cal E}_z  &=& 4 \pi \Sigma ~~ {\rm for} ~~ p_z \ge p_a ~~, ~~ {\cal E}_z = -4 \pi \Sigma ~~ {\rm for}~~ p_z \le -p_a, \nonumber \\
    {\rm and}~~~ {\cal E}_z &=& 0 ~~ {\rm for} ~~ -p_a \le p_z \le p_a. 
\end{eqnarray}
The momentum ``electric" field of \eqref{m-capacitor} is zero between the plates and non-zero outside the plates, which is the inverse of the standard capacitor \eqref{capacitor}. 

The reason for building our momentum gauge field capacitor system as the {\bf inverse} of the normal capacitor is due to the connection between the non-commutativity parameter, $\Theta _{\mu \nu}$ and the momentum gauge field tensor, $G_{\mu \nu}$, as given equations \eqref{nc-geo-1} and \eqref{x-com-1} {\it i.e.} $\Theta _{\mu \nu} = g G_{\mu \nu}$. We want to have a normal position-position commutator ({\it i.e.} $[X_\mu, X_\nu] =0$) for momenta near zero ({\it i.e.} for $-p_a \le p_z \le p_a$) but we want non-commutative space-time effects for large momenta {\it i.e.} we want $\Theta _{\mu \nu} \propto G _{\mu \nu} \ne 0$ for large momenta,  $|p_a| \le |p_z|$. This is different from the usual  non-commutative space-time approach where the non-commutative parameter is ``turned on" for all momentum. Here the non-commutativity, at least for the $\Theta _{0 i}$ components, is turned on only for $z$-momentum magnitude satisfying $|p_a| < |p_z|$. 

\subsection{Current sheet-type momentum magnetic field}

In this subsection we carry out a similar construction as in the preceding subsection, but for the space/space components of $\Theta _{\mu \nu}$ and $G_{\mu \nu}$. In this case the standard gauge field system we want to build a momentum gauge field analog of is two infinite plane sheet currents located at $z=\pm a$. These current sheets are symmetrically placed on the $z$-axis around $z=0$. The explicit surface currents are
\begin{equation}
    \label{sheet}
    {\vec K} = \pm J {\bf {\hat y}} ~~~{\rm at} ~~~ z= \mp a
\end{equation}
This leads a regular magnetic field of 
\begin{equation}
    \label{b-sheet}
    {\vec B}= 4 \pi J {\bf {\hat x}} ~~~ {\rm for} ~~ -a \le z \le a ~~~{\rm and} ~~~ {\vec B}= 0 ~~ {\rm for}~~ |a| \le |z|
\end{equation}
{\it i.e.} the magnetic field is a non-zero constant between the sheets and zero outside the sheets. 

The momentum gauge field analog of this is two momentum gauge field current sheets at the momentum planes, $p_z= \pm p_a$. These planes are symmetric around the origin through the $p_z$-axis. Explicitly the ``momentum" current sheets are
\begin{equation}
    \label{m-sheet}
     {\vec {\cal K}} = {\cal J} {\bf {\hat y}} ~~~{\rm at}~~~ p_z = \pm p_a
\end{equation}
Note that here we have the currents in the same direction, rather than opposite direction as for the regular gauge field current sheets of \eqref{sheet}. The reason for this is the same as for the momentum gauge field, capacitor-like system of the preceding subsection: we want the non-commutativity parameter to be zero for momentum in the range $-p_a \le p_z \le p_a$ and we want a non-zero non-commutativity parameter for momentum in the range $|p_a| \le |p_z|$. Putting this all together the momentum gauge field ``magnetic" field is 
\begin{eqnarray}
    \label{mb-sheet}
{\vec {\cal B}} &=& 4 \pi {\cal J} {\bf {\hat x}}  ~~{\rm for} ~~ p_a \le p_z ~~~~{\rm and} ~~~~{\vec {\cal B}} = - 4 \pi {\cal J} {\bf {\hat x}}  ~~{\rm for} ~~ p_z \le -p_a \nonumber \\
{\rm and} ~~ {\vec {\cal B}} &=& 0 ~~~~{\rm for}~~ -p_a \le p_z \le p_a  ~.
\end{eqnarray}
The momentum gauge ``magnetic" field is a non-zero, constant outside the current sheets and zero between the current sheets. This implies that the space/space non-commutativity parameter, $\Theta _{ij}$, is zero for momenta in the range $-p_a \le p_z \le p_a$, while for large magnitude momenta ({\it i.e.} $|p_a \le |p_z|$) the space/space component $\Theta _{yz} = g G_{yz} = g \epsilon_{yzx} {\cal B}_x = \pm g {\cal B}$ is a non-zero constant. Both this simple example and the example from the preceding subsection show that one can construct non-commutative space-times where the non-commutativity only ``turns" on at some large enough momentum, rather than being on all the time.

\section{Summary and conclusions}

In this paper we have studied the formulation of the gauge principle in momentum space, or energy-momentum space in the relativistic case. Instead of only starting with the momentum operator and introducing a covariant momentum as $p_i \to p_i - e A_i (x)$ we also considered the position operator and introduced a covariant position as $x_i \to x_i - g C_i (p)$. The preference for having only the covariant momentum and not the covariant position comes from the fact that in general one starts with a free Hamiltonian \eqref{H} which has only momentum dependence. However a more symmetric treatment, motivated by the fact that the QFT vacuum can be seen as a collection of oscillators, leads to a Hamiltonian of the form given in \eqref{H-sho} which then calls for both covariant momentum {\it and} covariant position. 

We presented several simple examples of this momentum formulation of the gauge principle, showing that one could construct momentum gauge field analogs to plane wave solutions, point charge solutions, and to Landau levels. All these examples are underpinned by a dual symmetry, exchange symmetry or reciprocity \cite{born1949} between momentum and position, namely  ${\hat x} \to {\hat p}$ and ${\hat p} \to -{\hat x}$ (or also ${\hat x} \to {\hat p}$ and ${\hat p} \to {\hat x}$) which then relates the regular gauge fields to the momentum gauge fields. A criticism of this momentum formulation of the gauge principle is whether or not it has any concrete physical application or use. In this regard we mention that the model presented here is similar to Born's reciprocity theory \cite{born1949} which Born had hoped would play a role in the theory of elementary particle. The new feature here is that our version of Born's reciprocity is dynamical since we have introduced momentum gauge fields \eqref{mom-gauge-field} and momentum field strength tensors \eqref{guv}. We will explore physical consequences of this idea in future work. 

One potentially interesting application of this momentum gauge theory is that it naturally lead to non-commutative geometry as given in equations \eqref{x-com} and \eqref{x-com-1}. This non-commutativity of space-time has been studied previously as a way to extend QFT \cite{Noncommutativefieldtheory}, as a way to test for extensions to QED \cite{NoncommutativeHydrogenatom}, and as a way to deal with the singularities of general relativity \cite{piero-2009}. The non-commutativity of these works rests on non-trivial space-time commutators of the form \eqref{nc-geo-1} where the non-commutativity parameter, $\Theta _{\mu \nu}$ is a constant. In our formulation, since the non-commutativity parameter is a momentum gauge field, field strength tensor, $g G_{\mu \nu}$, it can vary with momentum, since the momentum gauge field, $C_\mu$ can vary with momentum.  In section IV we constructed a very simple system, based on the infinite charge sheets and infinite current sheets of introductory E\&M, where the non-commutativity parameter, $g G_{\mu \nu}$, would only turn on when the magnitude of the momentum became large enough {\it i.e.} when the momentum satisfied $|p_a| \le |p_z|$ for some large, fixed $p_a$. This could have interesting consequences since one could have commutative space-time below $p_a$ that turns into non-commutative space-time above $p_a$.

These are examples to show that the non-commutativity parameters can be screened by ``charges" in the infrared or ultraviolet regions. This screening has been studied here by introducing external momentum currents, but they could also rise from quantum fluctuation, as in ordinary gauge theories where coupling constants are screened in  
the infrared ({\it i.e.} in QED) or ultraviolet ({\it i.e.} in QCD) regions. 
Another subject that could be studied is the possibility of coordinate gauge fields and momentum gauge field mixing and/or oscillating. This is suggested for example by the result that
minimally coupling both coordinate gauge fields and momentum gauge fields produces a very specific linear combination \eqref{Beff}  that couples to the angular momentum of matter. Thus after integrating out the matter, we should be left with coordinate gauge fields and momentum gauge field mixing and/or oscillating. There are also no obstacles to considering non-Abelian momentum gauge fields. Furthermore, a connection between momentum gauge fields and curved momentum space can be established, where the momentum gauge field appear from a higher dimensional curved momentum space from a Kaluza Klein mechanism \cite{KKmomentum}. This momentum space Kaluza-Klein approach could be further extended and could provide additional insights into higher dimensional theories.
Lastly the present authors have worked on other ways to modify the gauge principle with non-vector gauge fields \cite{dual-gauge-2,scalar-gauge, oldscalar-gauge} or by gauging a dual symmetry, \cite{dual-gauge}. Also the ``Curtright generalized gauge fields" presents yet another way to generalize the gauge principle \cite{curtrightgauge}. However, the present way of modifying the gauge principle that can have an additional symmetry principle underlying it namely the exchange symmetry or Born reciprocity where ${\hat x} \to {\hat p}$ and ${\hat p} \to -{\hat x}$ or ${\hat x} \to {\hat p}$ and ${\hat p} \to {\hat x}$ {\it i.e.} the role of momentum and position are exchanged . 

Motivated from the Born reciprocity, we may suspect that the simultaneous existence of momentum and coordinate gauge fields could have important consequences, for example in  
\cite{momentumandconfigurationgaugefieldstogether}
it is found that the simultaneous momentum-like Coulomb solution for momentum gauge fields, given by $C^0 = \frac{g}{p}$ and ${\vec C}= 0$,  together with the regular configuration space Coulomb solution, given by $A^0 = \frac{e}{r}$ and ${\vec A}=0$, can be related to the generation of an emergent spacetime.

\section{Acknowledgements}
We would like to thank FQXi - Foundational Questions Institute and  COST Action CA18108 - Quantum gravity phenomenology in the multi-messenger approach (QG-MM) for financial support and to Fabian Wagner for very interesting discussions.

\end{document}